\documentclass[article,preprint,groupedaddress]{revtex4}
\usepackage{epsfig}
\usepackage{graphicx}

\begin{document}
\title{A mystery of black-hole gravitational resonances}
%\title{\ \ \ \ \ \ \ \ \ \ \ A mystery of black-hole quasinormal resonances
%\newline
%[Comment on arXiv:1510.08159 by Zimmerman et. al.]}
\author{Shahar Hod}
\address{The Ruppin Academic Center, Emeq Hefer 40250, Israel}
\address{ }
\address{The Hadassah Institute, Jerusalem 91010, Israel}
\date{\today}

\begin{abstract}
\ \ \ More than three decades ago, Detweiler provided an analytical
formula for the gravitational resonant frequencies of
rapidly-rotating Kerr black holes. In the present work we shall
discuss an important discrepancy between the famous {\it analytical}
prediction of Detweiler and the recent {\it numerical} results of
Zimmerman et. al. In addition, we shall refute the claim that
recently appeared in the physics literature that the
Detweiler-Teukolsky-Press resonance equation for the characteristic
gravitational eigenfrequencies of rapidly-rotating Kerr black holes
is not valid in the regime of damped quasinormal resonances with
$\Im\omega/T_{\text{BH}}\gg1$ (here $\omega$ and $T_{\text{BH}}$ are
respectively the characteristic quasinormal resonant frequency of
the Kerr black hole and its Bekenstein-Hawking temperature). The
main goal of the present paper is to highlight and expose this
important {\it black-hole quasinormal mystery} (that is, the
intriguing discrepancy between the analytical and numerical results
regarding the gravitational quasinormal resonance spectra of
rapidly-rotating Kerr black holes).
\end{abstract}
\bigskip
\maketitle

\section{The black-hole quasinormal mystery}

One of the earliest and most influential
%most important
works on the quasinormal resonance spectra of black holes is
Detweiler's ``Black holes and gravitational waves. III. The resonant
frequencies of rotating holes" \cite{Det}. In this famous paper
\cite{Noteci}, he derived the characteristic resonance equation
\begin{equation}\label{Eq1}
-{{\Gamma(2i\delta)\Gamma(1+2i\delta)\Gamma(1/2+s-i\hat\omega-i\delta)\Gamma(1/2-s-i\hat\omega-i\delta)}\over
{\Gamma(-2i\delta)\Gamma(1-2i\delta)\Gamma(1/2+s-i\hat\omega+i\delta)\Gamma(1/2-s-i\hat\omega+i\delta)}}
=(-i\hat\omega\tau)^{2i\delta}
{{\Gamma(1/2+i\hat\omega+i\delta-4i\varpi/\tau)}\over{\Gamma(1/2+i\hat\omega-i\delta-4i\varpi/\tau)}}\
\end{equation}
for the quasinormal frequencies of near-extremal (rapidly-rotating)
Kerr black holes. Here \cite{Noteun,Notepar}
\begin{equation}\label{Eq2}
\tau\equiv 8\pi MT_{\text{BH}}\ \ \ ;\ \ \ \varpi\equiv
M(\omega-m\Omega_{\text{H}})\ \ \ ;\ \ \ \hat\omega\equiv 2\omega
r_+\ ,
\end{equation}
where
\begin{equation}\label{Eq3}
T_{BH}={{r_+-r_-}\over{8\pi Mr_+}}\ \ \ ; \ \ \
\Omega_{\text{H}}={{a}\over{2Mr_+}}
\end{equation}
are the Bekenstein-Hawking temperature and the angular velocity of
the rotating Kerr black hole, respectively. The parameters $\{s,m\}$
are the spin-weight and azimuthal harmonic index of the field mode
\cite{Teuk}, and $\delta$ is closely related to the
angular-eigenvalue of the angular Teukolsky equation
\cite{Teuk,Notese}.

Detweiler's resonance equation (\ref{Eq1}) is based on the earlier
analyzes of Teukolsky and Press \cite{Teuk} and Starobinsky and
Churilov \cite{StCh} who studied the scattering of massless spin-$s$
perturbation fields in the rotating Kerr black-hole spacetime in the
double limit $a/M\to1\ (T_{\text{BH}}\to 0)$ and $\omega\to
m\Omega_{\text{H}}$. These limits correspond to [see Eq.
(\ref{Eq2})]
\begin{equation}\label{Eq4}
\tau\to0\ \ \ \text{and}\ \ \ \varpi\to0\  .
\end{equation}

The rather complicated resonance equation (\ref{Eq1}) can be solved
{\it analytically} in two distinct regimes:
\newline
(1) In his original analysis \cite{Det}, Detweiler studied the
regime
\begin{equation}\label{Eq5}
{{\varpi}\over{\tau}}\gg1\ \ \ \text{with}\ \ \ \delta^2>0\  ,
\end{equation}
and obtained the black-hole eigenfrequencies
\begin{equation}\label{Eq6}
\varpi_n=-{{e^{\theta/2\delta}}\over{4m}}(\cos\phi+i\sin\phi)\times
e^{-\pi n/\delta}\ ,
\end{equation}
where the integer $n$ is the resonance parameter of the mode, and
\begin{eqnarray}\label{Eq7}
re^{i\theta}\equiv\Big[{{\Gamma(2i\delta)\over\Gamma(-2i\delta)}}\Big]^2
{{\Gamma(1/2+s-im-i\delta)\Gamma(1/2-s-im-i\delta)}\over
{\Gamma(1/2+s-im+i\delta)\Gamma(1/2-s-im+i\delta)}}\ \ \ ; \ \ \
\phi\equiv -{{1}\over{2\delta}}\ln r\  .
\end{eqnarray}
\newline
(2) On the other hand, in \cite{Hodm} (see also \cite{Hod1}) we have
analyzed the regime
\begin{equation}\label{Eq8}
{{\varpi}\over{\tau}}=O(1)\  ,
\end{equation}
and obtained the characteristic relation \cite{Hodm}
\begin{equation}\label{Eq9}
\Im\varpi_n=-2\pi T_{\text{BH}}(n+{1\over 2}+\Im\delta)
\end{equation}
for the {\it slowly} damped quasinormal resonances of the
rapidly-rotating (near-extremal, $T_{\text{BH}}\to0$) Kerr black
holes.

Recently, Yang. et. al. \cite{Yang} have used numerical techniques
to compute the quasinormal resonances of rapidly-rotating Kerr black
holes. They have found numerically \cite{Yang} that the slowly
damped quasinormal resonances of these near-extremal black holes are
descried extremely well by the analytical formula (\ref{Eq9}) of
\cite{Hodm,Hod1}. On the other hand, in their numerical study, Yang.
et. al. \cite{Yang} have found {\it no} trace of the black-hole
quasinormal resonances (\ref{Eq6}) predicted by Detweiler. This
discrepancy between the {\it analytical} prediction (\ref{Eq6}) of
\cite{Det} and the {\it numerical} results of \cite{Yang} is the
essence of the black-hole quasinormal mystery.

Most recently, Zimmerman et. al. \cite{Zimm} have claimed that the
discrepancy between Detweiler's {\it analytical} prediction
(\ref{Eq6}) and their {\it numerical} results \cite{Yang,Zimm} stems
from the fact that his resonance equation (\ref{Eq1}) is not valid
in the regime (\ref{Eq5}). In particular, they have claimed that the
standard matching procedure used in \cite{Teuk,StCh} to match the
near-horizon
\begin{equation}\label{Eq10}
x\equiv{{r-r_+}\over{r_+}}\ll1
\end{equation}
solution [see equation (A9) of \cite{Teuk}]
\begin{equation}\label{Eq11}
R={_2F_1}(-i\hat\omega+s+1/2+i\delta,-i\hat\omega+s+1/2-i\delta;1+s-4i\varpi/\tau;-x/\tau)\
\end{equation}
of the Teukolsky radial equation with the far-region
$x\gg\text{max}(\varpi,\tau)$ solution [see equation (A5) of
\cite{Teuk}] of the Teukolsky radial equation is invalid in the
regime (\ref{Eq5}) studied by Detweiler. Since Detweiler's analysis
is based on the matching procedure of \cite{Teuk,StCh}, Zimmerman
et. al. have claimed that Detweiler's analysis is also invalid in
the regime (\ref{Eq5}).

\section{The mystery is still unsolved}

In this paper we would like to point out that the assertion made in
Ref. \cite{Zimm} is actually erroneous. In particular, we shall show
below that the matching procedure of \cite{Teuk,StCh,Noteld} is
valid in the overlap region
\begin{equation}\label{Eq12}
\text{max}(\varpi,\tau)\ll x\ll 1\ .
\end{equation}

In their matching procedure, Teukolsky and Press \cite{Teuk} (see
also \cite{StCh}) use the identity [see Eq. 15.3.7 of \cite{Abram}]
\begin{eqnarray}\label{Eq13}
R={{\Gamma(1+s-4i\varpi/\tau)\Gamma(2i\delta)}
\over{\Gamma(-i\hat\omega+s+1/2+i\delta)\Gamma(1/2+i\hat\omega+i\delta-4i\varpi/\tau)}}
\Big({{x}\over{\tau}}\Big)^{i\hat\omega-s-1/2+i\delta} \\ \nonumber
\times{_2F_1}(-i\hat\omega+s+1/2-i\delta,-i\hat\omega+1/2-i\delta+4i\varpi/\tau;1-2i\delta;-\tau/x)+(\delta\to-\delta)\
\end{eqnarray}
for the near-horizon hypergeometric function (\ref{Eq11}) [The
notation $(\delta\to-\delta)$ in (\ref{Eq13}) means ``replace
$\delta$ by $-\delta$ in the preceding term."]. In order to perform
the matching procedure, Teukolsky and Press \cite{Teuk} (see also
\cite{StCh}) take the limit
\begin{equation}\label{Eq14}
{_2F_1}(-i\hat\omega+s+1/2{\mp}i\delta,-i\hat\omega+1/2{\mp}i\delta+4i\varpi/\tau;1{\mp}2i\delta;-\tau/x)\to
1\
\end{equation}
for the hypergeometric functions that appear in the expression
(\ref{Eq13}) of the radial eigenfunction $R$.

In Ref. \cite{Zimm} Zimmerman et. al. have recently claimed that the
limit (\ref{Eq14}) used in \cite{Teuk,StCh,Noteld} is invalid in the
regime (\ref{Eq5}) studied by Detweiler. To support their claim,
they plot (see Fig. 3 of \cite{Zimm}) the hypergeometric functions
of (\ref{Eq13}) in the regime
\begin{equation}\label{Eq15}
{{\varpi}\over{x}}=250\  .
%\varpi\sim1\ \ \ \text{with}\ \ \ x\sim 0.001\  .
\end{equation}
{\it Not} surprisingly, Zimmerman et. al. found that, in the regime
(\ref{Eq15}), the asymptotic behavior (\ref{Eq14}) used in the
matching procedure of \cite{Teuk,StCh,Noteld} is not valid. They
then concluded that the matching procedure of \cite{Teuk} is not
valid in the regime (\ref{Eq5}) studied by Detweiler.

However, here we would like to stress the fact that the analytical
arguments of Zimmerman et. al. (and, in particular, the data
presented in Fig. 3 of \cite{Zimm})
%is somewhat misleading.
are actually {\it irrelevant} for the discussion about the validity
of Detweiler's resonance condition (\ref{Eq1}). In particular, we
would like to emphasize the fact that the characteristic limiting
behavior
\begin{equation}\label{Eq16}
{_2F_1}(a,b;c;z)\to 1
\end{equation}
of the hypergeometric function is valid in the regime \cite{Noteah}
\begin{equation}\label{Eq17}
{{a\cdot b\cdot z}\over{c}}\ll1\  .
\end{equation}

Taking cognizance of the arguments $(a,b,c,z)$ of the hypergeometric
functions in (\ref{Eq13}), one realizes that, for moderate values of
the field azimuthal harmonic index $m$, the asymptotic behavior
(\ref{Eq14}) assumed in the matching procedure of
\cite{Teuk,StCh,Noteld} is valid in the regime [see Eqs.
(\ref{Eq13}) and (\ref{Eq17})]
\begin{equation}\label{Eq18}
\text{max}\Big(1,{{\varpi}\over{\tau}}\Big)\times
{{\tau}\over{x}}\ll1\ .
\end{equation}
In particular, one finds from (\ref{Eq18}) that, in the regime
$\varpi/\tau\gg1$ [see (\ref{Eq5})] studied by Detweiler, the
asymptotic behavior (\ref{Eq14}) used in \cite{Teuk,StCh,Noteld} is
valid for
\begin{equation}\label{Eq19}
{{\varpi}\over{x}}\ll1\  .
\end{equation}
This is certainly {\it not} the regime plotted in Fig. 3 of
\cite{Zimm} [see Eq. (\ref{Eq15})]. One therefore concludes that the
analytical arguments raised by Zimmerman et. al. (and, in
particular, the data presented in Fig. 3 of \cite{Zimm}) are simply
irrelevant
%\cite{Noteir}
for the discussion about the validity of the matching procedure used
in \cite{Teuk,StCh}!

Furthermore, taking cognizance of (\ref{Eq10}) and (\ref{Eq18}), one
concludes that the matching procedure of \cite{Teuk,StCh} is {\it
valid} in the overlap region
\begin{equation}\label{Eq20}
\text{max}\Big(1,{{\varpi}\over{\tau}}\Big)\times\tau\ll x\ll1\ .
\end{equation}
It is worth emphasizing again that Detweiler's analysis is based on
the matching procedure of \cite{Teuk,StCh}. As such, his resonance
equation (\ref{Eq1}) for the characteristic eigen-frequencies of
rapidly-rotating Kerr black holes is expected to be valid in the
regime (\ref{Eq20}) [and {\it not} in the regime (\ref{Eq15})
considered in Fig. 3 of \cite{Zimm}].

In Table \ref{Table1} we present the hypergeometric functions
(\ref{Eq13}) used in the matching procedure of \cite{Teuk,StCh}. We
display the values of these functions for the field mode $l=m=s=2$
with $n=2,3,4,5$ and $x=0.1$ [see (\ref{Eq10})]. This mode is
characterized by the angular eigenvalue $\delta=2.051$ \cite{PT},
which yields the Detweiler resonance spectrum [see Eqs. (\ref{Eq6})
and (\ref{Eq7})]
\begin{equation}\label{Eq21}
\varpi_{n}\simeq (0.162-0.035i)\times e^{-1.532n}\  .
\end{equation}

Inspection of the data presented in Table \ref{Table1} reveals that,
for resonant modes with $n\ge3$, the hypergeometric functions
${_2F_1}$ that appear in the expression (\ref{Eq13}) for the radial
eigenfunction $R$ are described extremely well by the asymptotic
behavior (\ref{Eq14}) as originally assumed in
\cite{Teuk,StCh,Noteld}. Our results therefore support the validity
of the matching procedure used by Teukolsky and Press \cite{Teuk}
and by Starobinsky and Churilov \cite{StCh}.

\begin{table}[htbp]
\centering
\begin{tabular}{|c|c|c|c|}
\hline
\ \ $\tau$\ \ & \ $10^{-4}$\ \ & \ \ $10^{-5}$\ \ & \ \ $10^{-6}$\ \ \\
\hline
\ \ $n=2$\ \ & \ \ $0.972-0.344i$\ \ \ & \ \ $0.972-0.348i$\ \ \ & \ \ $0.972-0.349i$\ \ \\
\ \ $$\ \ & \ \ $0.836-0.003i$\ \ \ & \ \ $0.836-0.003i$\ \ \ & \ \ $0.836-0.003i$\ \ \\
\hline
\ \ $n=3$\ \ & \ \ $1.003-0.071i$\ \ \ & \ \ $1.005-0.075i$\ \ \ & \ \ $1.005-0.075i$\ \ \\
\ \ $$\ \ & \ \ $0.961-0.001i$\ \ \ & \ \ $0.961-0.002i$\ \ \ & \ \ $0.961-0.002i$\ \ \\
\hline
\ \ $n=4$\ \ & \ \ $0.999-0.012i$\ \ \ & \ \ $1.001-0.016i$\ \ \ & \ \ $1.001-0.016i$\ \ \\
\ \ $$\ \ & \ \ $0.991-0.000i$\ \ \ & \ \ $0.991-0.000i$\ \ \ & \ \ $0.991-0.000i$\ \ \\
\hline
\ \ $n=5$\ \ & \ \ $0.998+0.001i$\ \ \ & \ \ $1.000-0.003i$\ \ \ & \ \ $1.000-0.003i$\ \ \\
\ \ $$\ \ & \ \ $0.998+0.000i$\ \ \ & \ \ $0.998-0.000i$\ \ \ & \ \ $0.998-0.000i$\ \ \\
\hline
\end{tabular}
\caption{The values of the hypergeometric functions (\ref{Eq13})
used in the matching procedure of \cite{Teuk,StCh} for the field
mode $l=m=s=2$ with $x=0.1$. The first row corresponds to the
hypergeometric function
${_2F_1}(-i\hat\omega+s+1/2-i\delta,-i\hat\omega+1/2-i\delta+4i\varpi/\tau;1-2i\delta;-\tau/x)$
in (\ref{Eq13}), whereas the second row corresponds to the
hypergeometric function
${_2F_1}(-i\hat\omega+s+1/2+i\delta,-i\hat\omega+1/2+i\delta+4i\varpi/\tau;1+2i\delta;-\tau/x)$
in (\ref{Eq13}). One finds that, for resonant modes with $n\ge3$,
these functions are described extremely well by the asymptotic
behavior (\ref{Eq14}).} \label{Table1}
\end{table}

\section{Summary}

Currently, we face a mystery which can be summarized as follows:
\newline
(1) Contrary to the claim made in \cite{Zimm}, Detweiler's resonance
equation (\ref{Eq1}) is valid in the regime (\ref{Eq5}).
\newline
(2) In his original {\it analytical} study, Detweiler has shown that
the resonance equation (\ref{Eq1}) predicts the existence of the
near-extremal Kerr black-hole quasinormal resonances (\ref{Eq6}) in
the regime (\ref{Eq5}).
\newline
(3) In their recent {\it numerical} study, Zimmerman et. al. have
found {\it no} trace of the black-hole quasinormal resonances
(\ref{Eq6}) predicted by Detweiler.

The main goal of the present work was to highlight the discrepancy
between the {\it analytical} prediction (\ref{Eq6}) of Detweiler and
the {\it numerical} results of Zimmerman et. al. We would like to
emphasize that currently we have no solution to this black-hole
quasinormal mystery. We believe, however, that it is important to
highlight this open problem. We hope, in particular, that the
present work would encourage researchers in the fields of black-hole
physics and general relativity to further explore this interesting
physical problem. Hopefully, future studies of the black-hole
quasinormal spectrum will shed light on this intriguing mystery.

\newpage
\bigskip
\noindent {\bf ACKNOWLEDGMENTS}
\bigskip

{\it In memory of Prof. Steven Detweiler}. This research is
supported by the Carmel Science Foundation. I would like to thank
Yael Oren, Arbel M. Ongo, Ayelet B. Lata, and Alona B. Tea for
stimulating discussions.


\begin{thebibliography}{99}
\bibitem{Noteci} To date, this paper \cite{Det} has been cited more than 130
times.
\bibitem{Noteun} We use natural units in which $G=c=\hbar=1$.
\bibitem{Notepar} Here $M, a\equiv J/M$, and $r_{\pm}=M+(M^2-a^2)^{1/2}$ are the mass,
angular momentum per unit mass, and horizon radii of the rotating
Kerr black hole, respectively.
\bibitem{Noteld} It is worth emphasizing again that, as originally shown by
Detweiler, the matching procedure used in \cite{Teuk,StCh} yields
the quasinormal resonance condition (\ref{Eq1}).
\end{thebibliography}
\end{document}